# Application of Liquid Rank Reputation System for Content Recommendation

Abhishek Saxena
*Mathematics and Mechanics Department*
*Novosibirsk State University*
Novosibirsk, Russia
abhishek.saxena.mnnit@gmail.com

Anton Kolonin
*Mathematics and Mechanics Department*
*Novosibirsk State University*
Novosibirsk, Russia
akolonin@gmail.com

***Abstract*—An effective content recommendation on social media platforms should be able to benefit both creators to earn fair compensation and consumers to enjoy really relevant, interesting, and personalized content. In this paper, we propose a model to implement the liquid democracy principle for the content recommendation system. It uses a personalized content recommendation model based on reputation ranking system to encourage personal interests driven recommendation. Moreover, the personalization factors to an end users' higher-order friends on the social network (initial input Twitter channels in our case study) to improve the accuracy and diversity of recommendation results. This paper analyzes the dataset based on cryptocurrency news on Twitter to find the opinion leader using the liquid rank reputation system. This paper deals with the tier-2 implementation of a liquid rank in a content recommendation model. This model can be also used as an additional layer in the other recommendation systems. The paper proposes the implementation, challenges, and future scope of the liquid rank reputation model.**

## I. Introduction

The most discussed problem in recommender systems is Mathew's effect on social commerce [1]. Another major and dangerous problem is skewing of results eg. using a large pool of bots working against the social media recommendations or manipulating the recommender systems by skewed data inputs eg. Covid misinformation through bots on Twitter [2]. Along with challenges to validate the validators in a system always remains the primary concern. To eliminate the weaknesses associated with the existing algorithm, a reputation system, which effectively expresses the idea of the so-called "liquid democracy" algorithm [3] has been introduced.

When applying to modern social media platforms, such as Facebook, Twitter, and Youtube an effective content recommendation algorithm should be able account for both content creators and content consumers. New content creators should be getting enough exposure and consumers should get really interesting, relevant and personalized content. Along with that accurate recommendations can improve the consumer's experience but it is regarded as a healthier content recommendation system when it encourages individuals, especially small content creators, to share their creative content.

One possible way is to use mechanism design approaches in the reputation algorithm to give a recommendation based on the higher-order friends of the particular person or peer-to-peer recommendation [4].

In this paper, we will see the implementation of the liquid democracy model in the recommendation model above through the liquid rank reputation system [3]. This has been achieved by working on a cryptocurrency dataset taken from Twitter to find the opinion leader or most important/effective/influential Twitter channel/personal in that time frame.

The implementation would be at tier-2 level i.e. the recommendation based on the higher-order friends/initiations would be rated by the content user. This is implemented by selecting the initial channels (cryptocurrency-related) from Twitter as a personal recommendation. Thus applying liquid ranking resulted in recommending channels that have better ratings and wider visibility for quality content channels from the dataset (even if they are new or less rated in numbers) Similar work done of basic formulation of the model based on the higher-order friends for content recommendations [5].

### A. Why Liquid Rank Reputation System?

The purpose of the reputation system is to support a transparent and reliable ecosystem, and unbiased content management. A reliable solution is important for peer-to-peer systems, where each node can communicate with every other node in the network [6]. A reputation system [3] is that it leverages recognition over every user on the social platform. This means that fame is more important than possession. Reputation rating works as calculating of reputation calculated for each node based on its performance over a social platform, communication channel, financial transactions, or any other platform from which a reputation over a greater period can be fetched. This makes the reputation system as the involvement of every node on the network for the verifier.

Section II presents the Methodology, Section III Results. Section IV Future work and scope and Section V Conclusion.

## II. Methodology

### A. Implementation of Liquid Democracy in Recommendation Model using Liquid Rank Reputation algorithm.

The following design of reputation calculations is based on previous works on reputation, also known as "social capital" or "karma" [3]. The reputation system tries to implement liquid democracy on cryptocurrency dataset scraped from Twitter to find the opinion leader in the given time frame.

Dataset from the Twitter network contained 37615 public posts ("tweets") across 39 well-known cryptocurrency popular feeds (channels). These channels serve as the initial input to the reputation system for content/channel recommendation or the opinion leader for the cryptocurrency domain. Hence, initial channels serving as inputs can be considered as higher-order friends on a social platform.

In this paper, the analysis is done on a Twitter dataset to find the opinion leader by only mentions and by using a liquid rank reputation system. This can be considered as a base content recommender system. One ranking for opinion leader as been found with just mentions (Fig. 1) and the other with a liquid rank reputation system (Fig. 2).

The paper makes the use case simple, also considering the restraints of the dataset (not getting all the metrics like followers, likes, etc) a simple computational model of reputation [2] is calculated as

$$Rj = \sum(Ri * \ Vijt) \tag{1}$$

where Vijt here is an implicit rating (positive) as the number of mentions by each channel for node j (Twitter channel) being rated, node i supplying the rating and time t. Reputation is calculated for channels, but it is assumed that each channel plays a multi-agent social network role for each user account. Nodes during their operation, so that their artifacts can be indirectly rated mentioned on Twitter network cryptocurrency feed by the channels which were selected as initial inputs (higher-order friends) to fetch the dataset in the given time frame.

A liquid ranking algorithm is developed to work as a hybrid feature for content recommendation (a dynamic recommendation model). Each rank has been calculated by the individuals on a platform, using mentions in the Twitter dataset by the channels (it can be based on either of followers, likes, comments, subscribers, etc also).

The algorithm uses normalization after each cycle of reputation scoring and then starts the cycle again till the reputation change is negligible (<0.0001). The code library for all the algorithms and analysis with results can be accessed at https://github.com/xenvik/Recommendation-Model/tree/main.

Finally, the top 50 channels recommended on both scales by mentions and by liquid ranking were rated on a 0-2 scale (for qualitative analysis). Rating for qualitative analysis was done manually for the top 50 channels considering the influence of each channel in the cryptocurrency domain. Number Tweets per day, Tweets related to cryptocurrencies, and "retweets" by the people and the number of followings were the main metrics to provide the ratings. In the given time frame of the dataset, 0 is given to channels that have the least influence and 2 to the most influence on the community.

# III. Results

## A. *Mentions versus Liquid Ranking*

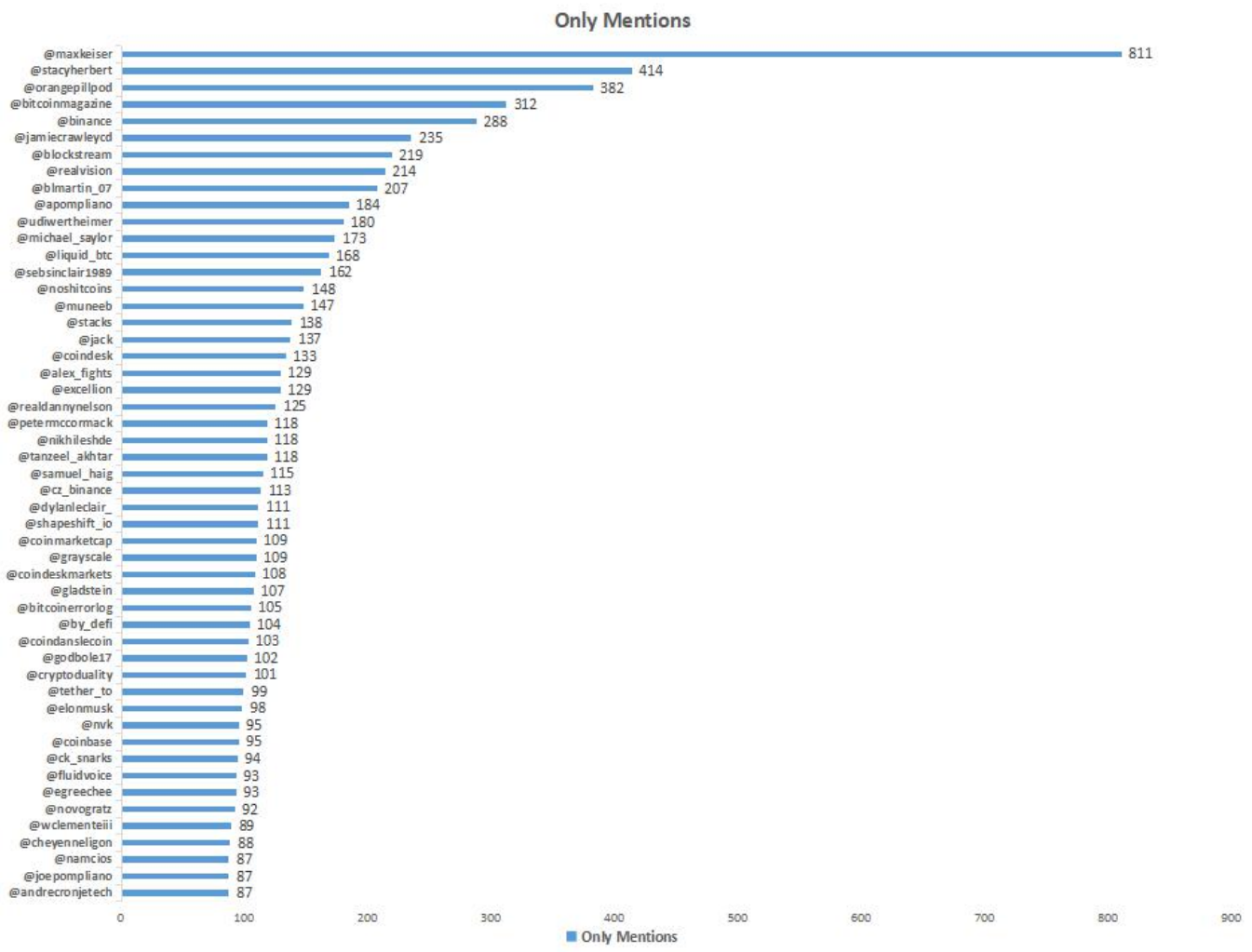

Fig. 1. Ranking based on Mentions in Data (A)

It shows that the top-ranking mentions in the list of ranking by just mentions are the Twitter channels which are of course very influential and have large followings. These are the channels which apart from large followings, may also come from direct relations with large channels eg. Channel Binance has maximum mentions for channel "maxkeiser" in the given time frame, the channel Binance in itself has a large following. This metric does not take care of the importance of mentions by more number of channels instead just the count of mentions.

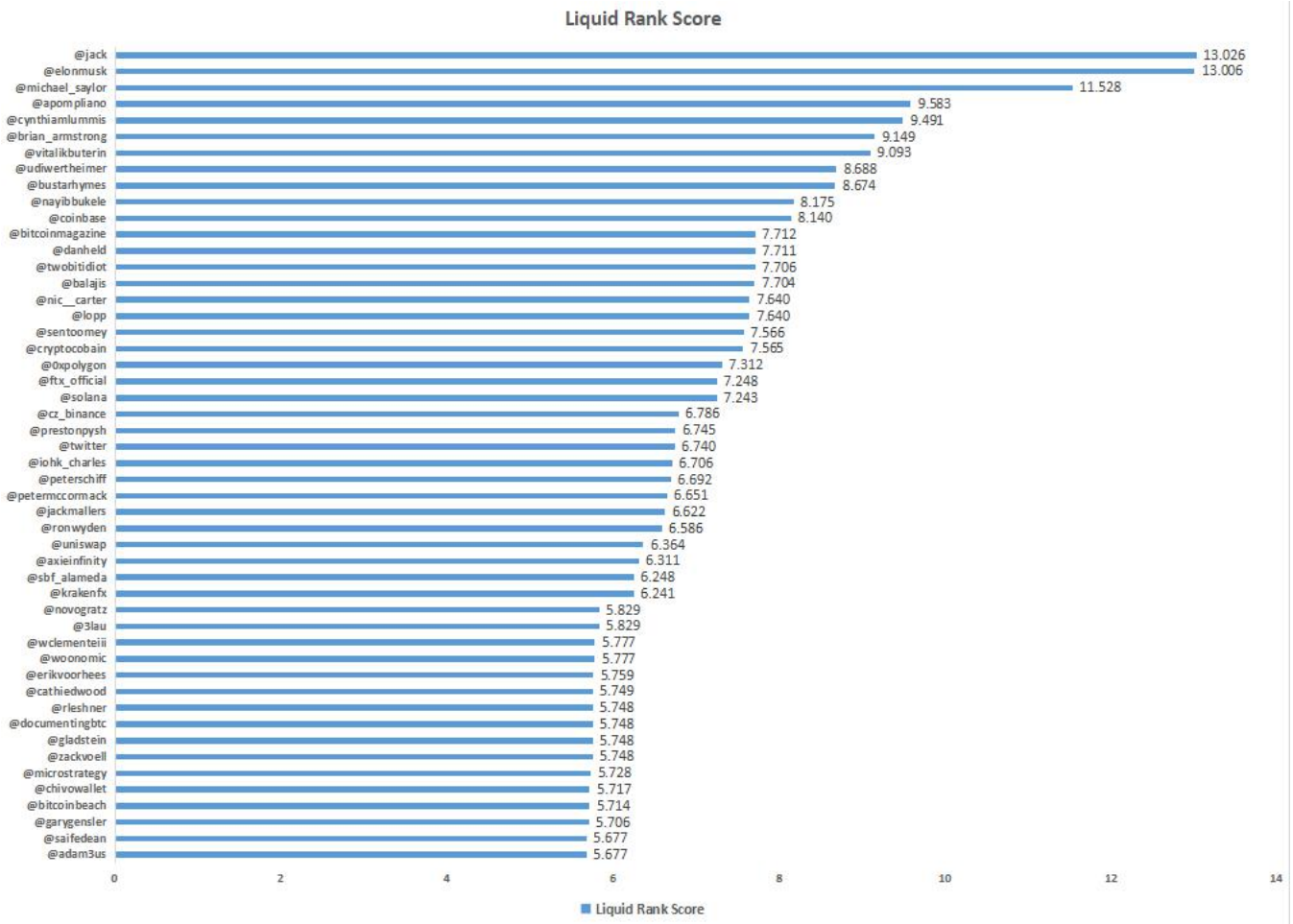

Fig. 2. Liquid Ranking from Data (B)

It shows that the top leaders in the Liquid Ranking are not only the channels which were mentioned a lot but also are the ones mentioned across different channels more. Liquid ranking using reputation score have normalized the impact of mentions by count with mentions across channels, hence we see that channel maxkeiser even after having the maximum number of mentions is not even in the top 50 ranking list by liquid reputation score.

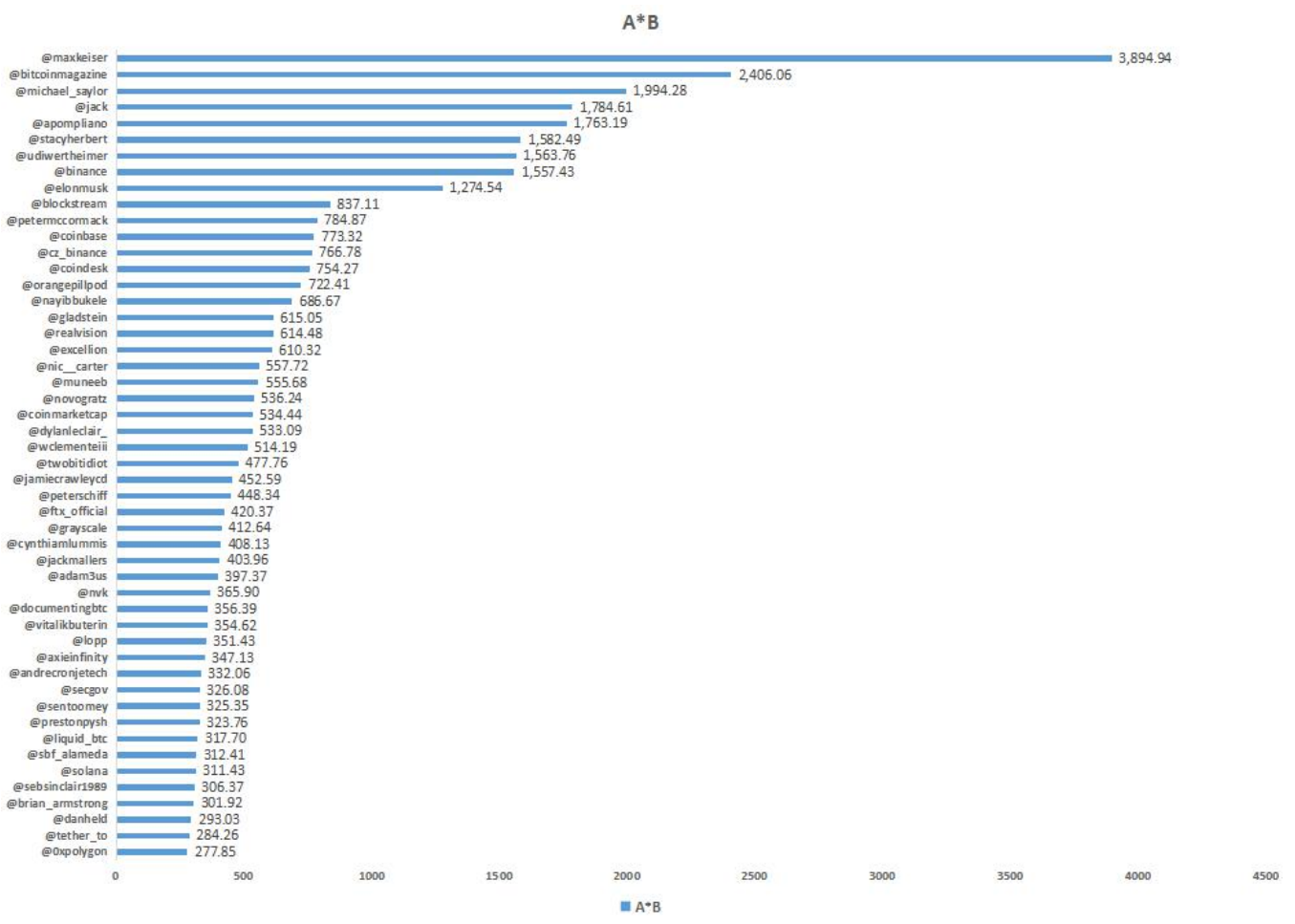

Fig. 3. Ranking based on A*B

Overall product of the two metrics as shown in Fig. 3 has an oversimplified impact of rank by mentions on the final results, as we see above the top place is again taken up by channel maxkeiser. One more thing that can be fetched in the current stage is those top leaders in the liquid ranking are not essentially very popular channels but were somehow popular in that time frame of data fetching.

## B. *Qualitative Analysis on the Results*

Qualitative analysis was done for the results based on ranking by mentions and liquid rank reputation system on three metrics decision support, average precision method, and mean reciprocal ranking method (Fig. 4-8) to analyze how liquid ranking based on mentions is effective versus ranking by just mentions.

The top 50 list from both rankings by mentions and by reputation score have been rated as stated in methodology based on relevance and impact on cryptocurrency domain on a scale of 0 to 2. These three metrics have been applied to analyze the top 50 ranks in both rankings for the maximum relevant recommendations by both systems. The rankings with qualitative score 2 have been considered relevant and others with scores 0 and 1 are considered irrelevant.

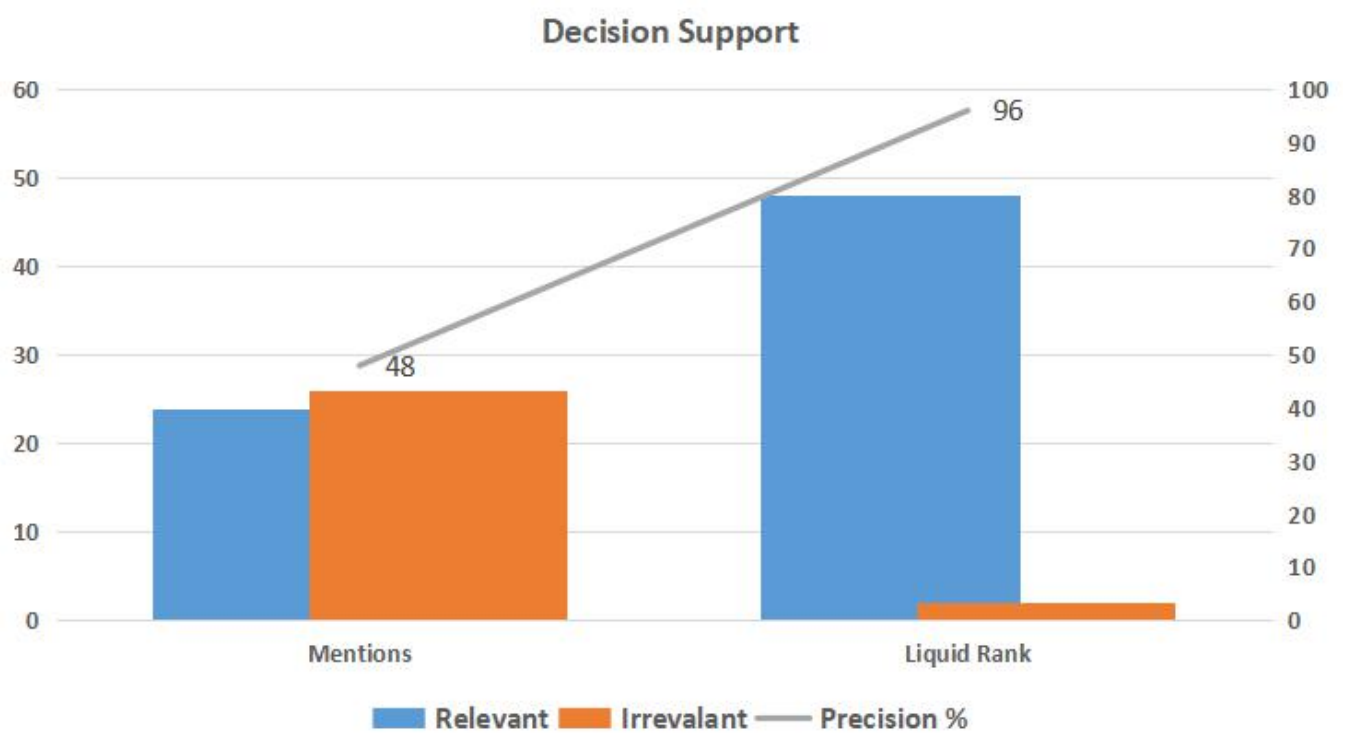

Fig. 4. Decision Support (Precision on Results)

The decision support method used above as shown in Fig. 4 is based on the precision metric i.e. the number of relevant quantities are considered concerning the overall quantities. Hence, we see that the ranking by mentions has almost 48% precision score when all the top 50 ranks are considered in comparison with liquid ranking which has a massive 96% precision results, just double of ranks by mentions. It can be considered that the channels in the top 50 list of liquid ranking are most relevant and have a high impact on the cryptocurrency domain, especially in the time frame of the dataset.

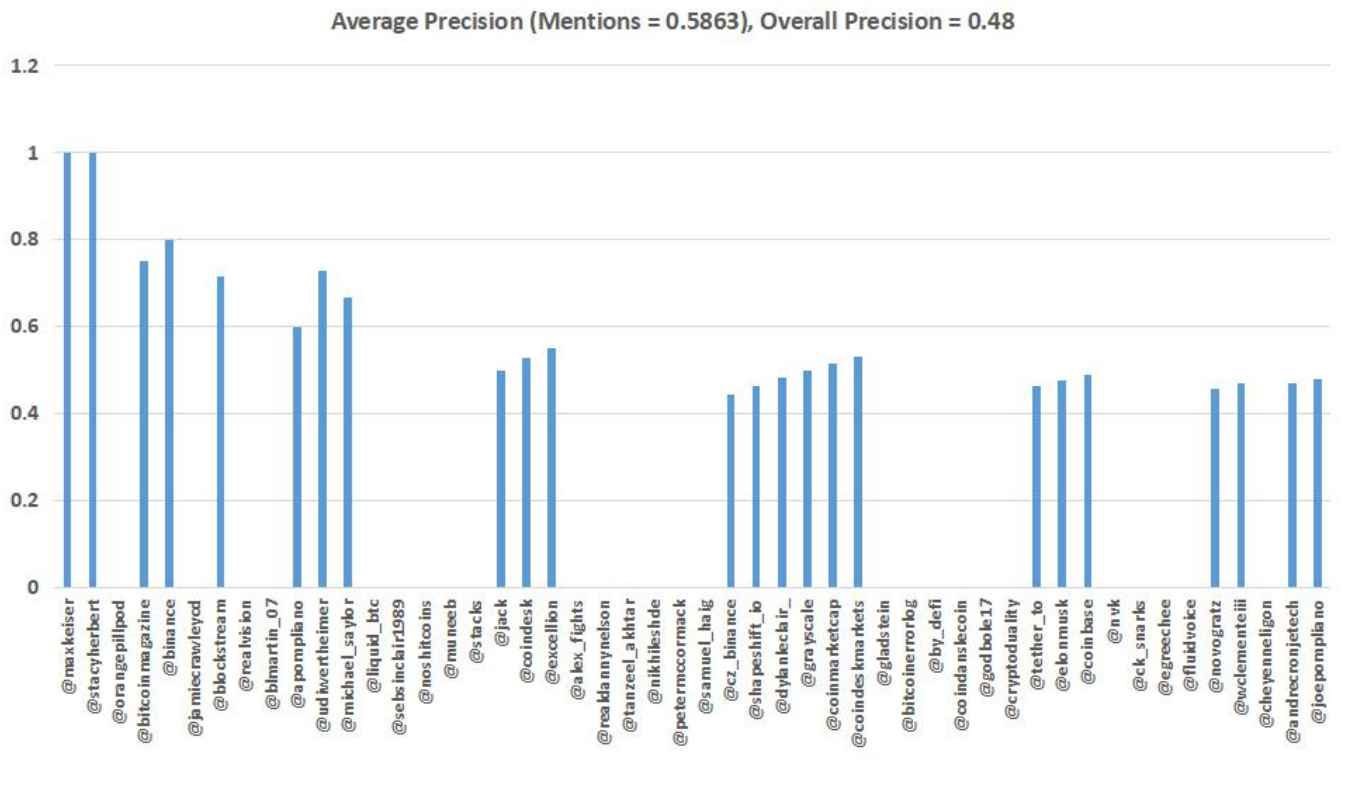

Fig. 5. Average Precision based on Mentions Ranking

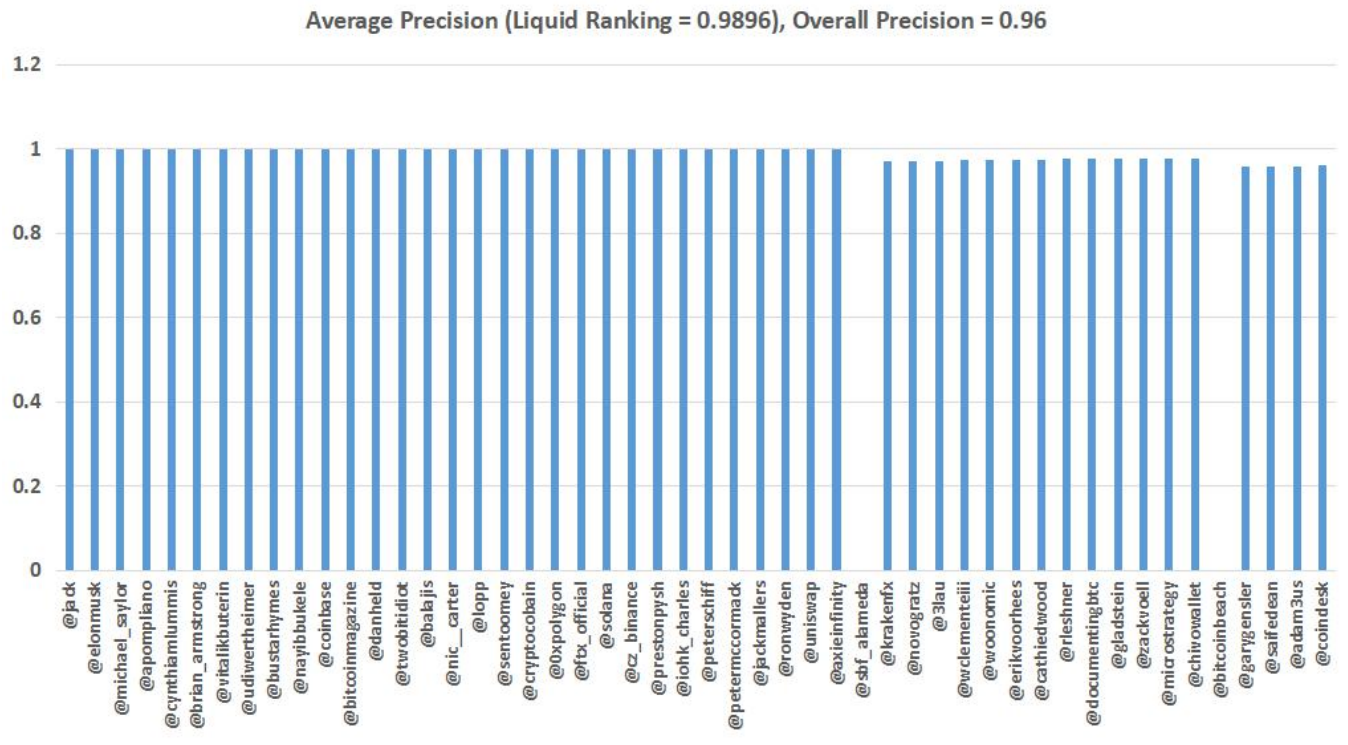

Fig. 6. Average Precision based on Liquid Ranking

The second metric was the average precision method on the results based on both rankings. The results are shown in Fig. 5 and 6. Average precision gives a strong metric for checking the relevancy of the topmost rankings and overall rankings too in our case. The relevant ranks have been re-ranked. Thus we see average precision in mentions ranking is 0.5863, showing that top ranks in the top 50 list by mentions have average precision around 58%, but overall precision is 48%, which implies there are more irrelevant ranks in lower ranks than the top ranks. Whereas, average precision in liquid ranking is 0.9896 and 0.96, which implies the same that the irrelevant ranks are further in lower ranks and overall the system by reputation ranking have very high precision.

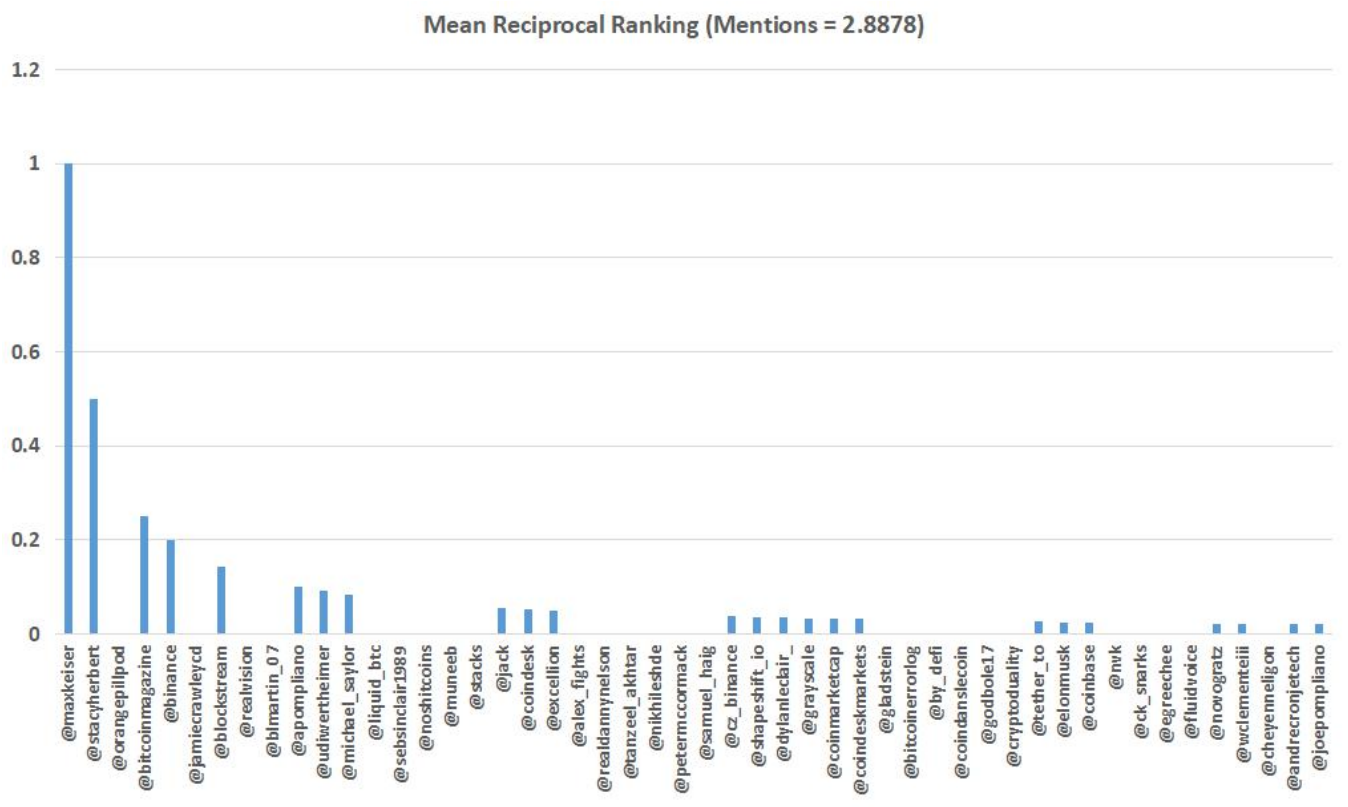

Fig. 7. Mean Reciprocal Ranking based on Mentions

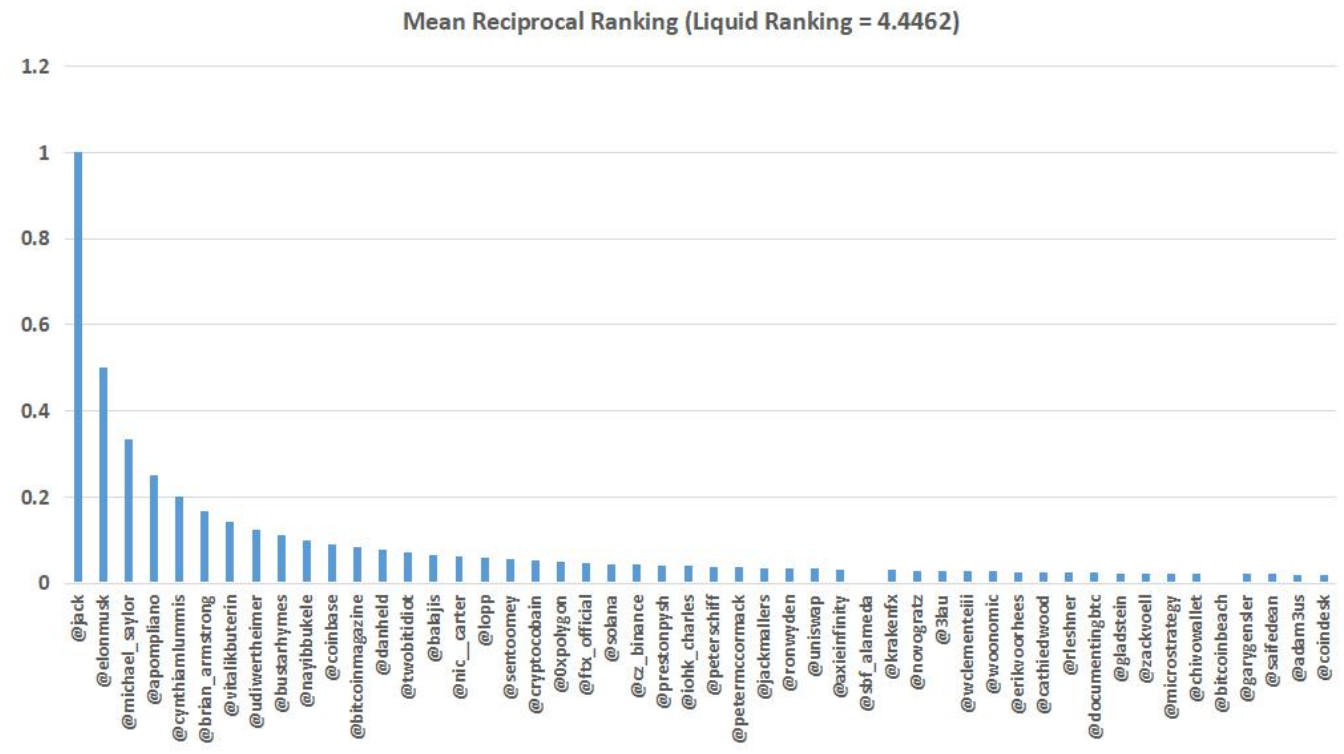

Fig. 8. Mean Reciprocal Ranking based on Liquid Ranking

The last qualitative analysis metric is mean reciprocal ranking. The results are shown in Fig. 7 and 8. Mean reciprocal rank focuses on where is the first relevant item in the recommended list. The rating for a list with the first relevant item at its third position will be greater than for a list with the first relevant item at 4th position. Since we already know that the relevant ratings in liquid rankings are 96%, the overall mean reciprocal ranking score is 4.4462 much higher than that of ranking by mentions 2.8878. The curve for liquid ranking is smoothly decreasing showing all the relevant ranks comprising a higher score, whereas ranking by mentions has many blanks and a discontinuous curve.

## IV. Future work and Scope

Liquid Ranking can directly be implemented in the account of social media channels or an online store where the cold start would be handled by the reputation of our friends and people whom we already follow. Later this reputation would ripple to more recommendations, starting from our social circle.

Liquid Ranking on multiple attributes can make not only personalized recommendations much better but also will help new content providers or say here in the case less influential channels also to be recommended (who are producing good work and talked about in our social circle, not viral but surely liked and effective content)

Reputation systems have a very capable foundation to build the ecosystem which could change the way how entertainment, news, or any social media platform offers recommendations. If considered now in the digital society, with the power of the reputation system, the whole imbalance can be controlled in the coming future.

Personalized, e.g. considering Twitter data, personal connections' following can be used as a base of liquid ranking to get recommendations. Dynamic, each feature used as a base of liquid ranking can be seen as a timestamped transaction, hence the recommendation can be dynamic.

The flexibility of such a reputation scoring system helps in binding the complexity level according to the environment required. It can be a social platform reputation system, a recommendation authorization, or a more closed centralized working system where only certain validated nodes are needed.

Moreover, it can be based on calculation performance required at certain periods such as hourly, daily, weekly, or monthly levels. (We can add further levels for features like sharing, promoting, etc using the reputation depth model in the liquid democracy).

## V. Conclusion

The present paper showed the implementation of a liquid rank reputation system for liquid democracy model on the Twitter cryptocurrency dataset to find the opinion leader. To make the simpler use case and practicability, considering the scope of the metrics available and time, the analysis was done on the most transparent metric, mentions of the channels during a given time frame. Qualitative analysis methods used were decision support, average precision method, and mean reciprocal rank.

Henceforth the analysis, we see that on all the three qualitative metrics liquid rank reputation system is much more relevant and better in terms of providing a recommendation or opinion leader in our case in the given time frame within the boundaries of the dataset. A liquid rank reputation system can be used with multiple metrics to make it further a strong and varied recommendation model.

The results of this study can help provide giving a balance to the content creator environment, along with providing better recommendations for the content user.

The liquid rank reputation system has high commercial value as it can be hybridized as a layer, for personalized and dynamic recommendations with already present recommendation models to further enrich them with more liberties and better choices for the users' taste on any selected metrics. Not only this, it can further promote and help small channels providing good content to come up with recommendations and grow as per their good work.